\documentclass[aps,twocolumn,showpacs,preprintnumbers,floatfix]{revtex4}
\usepackage{graphicx}
\usepackage{epsfig}
\usepackage{amsmath}
\usepackage{dcolumn}% Align table columns on decimal point
\usepackage{bm}% bold math
\usepackage[colorlinks=true]{hyperref}

\def\be{\begin{equation}}
\def\ee{\end{equation}}
\def\bea{\begin{eqnarray}}
\def\eea{\end{eqnarray}}
\def\bsp{\be\begin{split}}
\def\la{\langle}
\def\ra{\rangle}
\def\dag{\dagger}

\begin{document}
\title{Wave functions of $SU(3)$ pure gauge glueballs on the lattice}
\author{\small Jian Liang,$^{1,}$\footnote{liangjian@ihep.ac.cn} Ying Chen,$^{1,}$\footnote{cheny@ihep.ac.cn}
Wei-Feng Chiu,$^{1}$ Long-Cheng Gui,$^{2}$ Ming Gong,$^{1}$ and Zhaofeng Liu $^{1}$}
\affiliation{\small ${}^1$ Institute of High Energy
Physics, Chinese Academy of Sciences, Beijing 100049, China\\
${}^2$ Department of Physics, Hunan Normal University, Changsha 410081, China}
\begin{abstract}
{
  The Bethe-Salpeter wave functions of $SU(3)$ pure gauge glueballs are
  revisited in this study. The ground and the first excited states of the scalar
  and tensor glueballs are identified unambiguously through the variational method.
  We calculate the wave functions in the Coulomb gauge and use two lattice spacings
  to check the discretization artifacts. For the ground states, the radial wave functions are
  approximately Gaussian and the size of the tensor glueball is roughly twice as
  large as that of the scalar glueball. For the first excited states, the radial nodes
  are clearly observed for both the scalar and the tensor glueballs, such that they can be
  interpreted as the first radial excitations. These observations may shed light
  on the theoretical understanding of the inner structure of glueballs.
}
\end{abstract}
\maketitle
\section{introduction}
In quantum chromodynamics (QCD), gluons have strong interactions with each
other and can form a new type of hadron, the glueball, which is distinct from
the conventional $q\bar{q}$ mesons in the picture of the quark model.
However, there is not yet a reliable theoretical
description of the intrinsic degrees of freedom of glueballs. Phenomenologically,
two-gluon glueballs are the simplest low-energy color singlet systems made up of gluons,
whose $J^{PC}$ quantum numbers are expected to be $0^{++}$, $2^{++}$, $0^{-+}$, etc.
($1^{++}$ quantum number does not appear for two-gluon glueballs if the constituent gluons are massless).
These arguments are in qualitative agreement with the pattern of the glueball spectrum obtained by lattice QCD calculations~\cite{Morningstar:1997er, Morningstar:1999cu,
Chen:2006iy, Gregory:2012cr}, where the lowest-lying states are
$0^{++}$, $2^{++}$, $0^{-+}$ glueballs in the order of mass from low to high.
As for the gluon dynamics inside glueballs, there are various effective approaches such as the bag
model~\cite{Carlson:1983}, QCD in Coulomb
gauge~\cite{Szczepaniak:1996, Szczepaniak:2003}, and potential
models~\cite{Fritzsch:1981, Barnes:1981, Cornwall:1983}. The potential
models depict a glueball as a bound state of two or more constituent
gluons through a confining interacting potential. Even though still
controversial, the potential models with proper theoretical assumptions can reproduce the glueball spectrum
from lattice QCD.

Apart from the spectrum, some other static properties of glueballs,
such as the Bethe-Salpeter (BS) wave functions, can also be
investigated through lattice QCD studies, from which one can infer
some qualitative information on the sizes and the spatial profiles of
glueballs. For a glueball dominated by a two-gluon component, the BS
wave function is defined through the matrix element of a
two-gluon operator $A_\mu(x)A_\nu(y)$ between the glueball state and
the vacuum, which can be derived from the relevant two-point function calculated in lattice QCD.
This kind of wave function may reflect
the spatial structure of a hadron to some extent. Since the two-gluon
operator $A_\mu(x)A_\nu(y)$ is obviously gauge variant, the related
calculation should be performed in a fixed gauge, for example, the
Coulomb gauge. The pioneering study on this topic was carried out
for the $SU(2)$ pure gauge glueballs~\cite{deForcrand:1991kc}. A
similar study on $SU(3)$ glueballs can be found in Ref.~\cite{Loan:2006hw}.

There are actually phenomenological studies~\cite{Buisseret:2007cx,Buisseret:2009dk} on the gluon
dynamics within glueballs by the use of the BS wave functions obtained from lattice QCD
calculations, where the positive C-parity states are taken as mainly two-gluon states and the
interacting potential between the constituent gluons is extracted. However, the previous lattice
studies focus only on ground states and the results are not precise enough, thus a more scrutinized
analysis is desired for the BS wave functions of both the ground and the excited states. This is
exactly the major goal of this work. It is known that the key point in the calculation of the
matrix elements mentioned above is to identify the glueball states unambiguously. In order for
this, we adopt the sophisticated technique used in the glueball spectrum
study~\cite{Morningstar:1997er, Morningstar:1999cu, Chen:2006iy}. This technique implements the
variational method based on large operator sets for glueballs, through which the ground and the
first excited states can be well determined.  Finally, the BS wave functions of all the glueball
states are derived to a high precision. We hope our results can provide more information to the
understanding of the nature of glueballs.

This paper is organized as follows. Section{}~\ref{Two-gluon operators} gives a detailed
introduction to the definition of two-gluon operators on the lattice. Section{}~\ref{numerical
details} contains the numerical details in calculating the BS wave functions. In
Section{}~\ref{numerical details} we also discuss the parameterizations of the wave functions and
the corresponding physical implications. A brief summary is given in Section{}~\ref{summary}.

\section{Two-gluon operators}
\label{Two-gluon operators} It is known that the Bethe-Salpeter equation~\cite{Salpeter:1951}
provides a relativistic description of a two-body system where the relativistic Bethe-Salpeter (BS)
wave function of a bound state can then be defined. For a two-gluon glueball state $|G\rangle$ in
its rest frame, the BS wave function can be expressed as \cite{deForcrand:1991kc}: \be
\label{wavefunc} \Phi_G(r)=\la 0|\sum_{\mu,\nu}
\alpha(\mu,\nu)\sum_{|\vec{r}|=r,\vec{x}}Y_{lm}(\hat{r})A_\mu(\vec{x})A_\nu(\vec{x}+\vec{r})|G\ra,
\ee where $\hat{r}$ stands for the spatial orientation of $\vec{r}$, the summations on $(\mu,\nu)$
and $|\vec{r}|$ guarantee the right quantum number of the state $|G\rangle$, and the gauge field
$A_{\mu}$ acts as the creation operator for a gluon. Since the fundamental gluonic variables on the
lattice are the gauge links $U_{\mu}(x)$ instead of $A_{\mu}(x)$, one has to build the lattice
counterpart of the two-gluon operator $O_{A,\mu\nu}({x},\vec{r})=A_\mu({x})A_\nu({x}+\vec{r})$
through gauge links $U_\mu(x)$.

The gauge field $A_\mu(x)$ can be directly defined as
\begin{equation}
A_\mu(x)\sim  [U_\mu(x)-U_\mu(x)^\dagger](1 + O(a^2))
\end{equation}
by the use of the relations
\begin{equation}
U_\mu(x)=e^{iagA_\mu(x)},~~~U^\dag_\mu(x)=e^{-iagA_\mu(x)},
\end{equation}
and the classical small-$a$ expansion of $U_{\mu}(x)$.
The two-gluon operator through this way is denoted as $O_{A}^{1}$ in this work.
In Ref.~\cite{deForcrand:1991kc} the authors propose an alternative definition of the
two-gluon operators (denoted as $O_{A}^{2}$)
 \begin{equation}
  \begin{split}
    \mathcal{O}^2_{A,\mu\nu}(x,\vec{r})&=ReTr[U^\dag_\mu(x)U_\nu(x+\vec{r})]\nonumber\\
                                       &-\frac{1}{3}ReTr[U_\mu(x)]\cdot ReTr[U_\nu(x+\vec{r})]\\
                                       &=(ag)^2A_\mu(x)A_\nu(x+\vec{r})+\mathcal{O}(a^4),
  \end{split}
\end{equation}
and claim that this definition reduces the possible mixing with the flux states.
Although flux states do not appear in this work (see below), we also carry out the
relevant calculation using $O_{A}^{2}$ to check the possible lattice artifacts
owing to the different definitions of the two-gluon operators.

However, the classical expansion of $U_\mu(x)$ is not very justified due to the tadpole diagrams if
the quantum effects are considered. In view of this fact, we propose an alternative way to define
$A_\mu(x)$ from $U_\mu(x)$ through a non-linear derivation,
\begin{equation}\label{define_A}
A_\mu(x)\sim {\rm log}[U_\mu(x)].
\end{equation}
For any $U_\mu(x)$, which is an element of the $SU(3)$ group, there exists an unitary matrix $R$
satisfying \be R^\dag U R ={\rm diag}(\lambda_1,\lambda_2,\lambda_3), \ee with $\lambda_i$,
$i=1,2,3$ being the 3 eigenvalues of the matrix $U$. Thus Eq.~(\ref{define_A}) can be solved
exactly as
\begin{equation}
A_\mu(x)\sim R\cdot {\rm diag}\left({\rm log}[\lambda_1],\rm{log}[\lambda_2],{\rm
log}[\lambda_3]\right)\cdot R^\dag.
\end{equation}
The subsequent two-gluon operators are called $\mathcal{O}_A^3$ in this work.

The spatial symmetry group on a lattice is the discrete octahedral point group $O$ instead of the
$SO(3)$ group in the continuum limit, whose irreducible representations are $R=A_1, A_2, E, T_1,
T_2$. Although the correspondences of these irreducible representations to the angular momenta $J$
are not unique, it is always conjectured that the lowest states in each symmetry channel $R$
correspond to the states with lowest $J$'s in the continuum. To be specific, the quantum number of
the scalar glueball ($J^{PC}=0^{++}$) is realized by the $A_1^{++}$ irreducible representation on
the lattice, and that of the tensor glueball is given by both $E^{++}$ and $T_2^{++}$. The
two-gluon operators belong to the above irreducible representatives are expressed explicitly as \be
\begin{split}
\mathcal{O}_A^{A_1^{++}}(r)&=\sum_{|\vec{r}|=r,\vec{x}}\sum_{i=1}^{3} A_i(\vec{x}+\vec{r}) A_i(\vec{x}),\\
\mathcal{O}_A^{E^{++},1}(r)&=\sum_{|\vec{r}|=r,\vec{x}}\frac{1}{\sqrt{2}}\left( A_1(\vec{x}+\vec{r}) A_1(\vec{x})
                     -A_2(\vec{x}+\vec{r}) A_2(\vec{x})\right),\\
\mathcal{O}_A^{E^{++},2}(r)&=\sum_{|\vec{r}|=r,\vec{x}}\frac{1}{\sqrt{6}}\left( 2 A_3(\vec{x}+\vec{r}) A_3(\vec{x})
                     -A_1(\vec{x}+\vec{r}) A_1(\vec{x})\right.\\
&\left. -A_2(\vec{x}+\vec{r}) A_2(\vec{x})\right),\\
\mathcal{O}_A^{T_2^{++},1}(r)&=\sum_{|\vec{r}|=r,\vec{x}}\frac{1}{\sqrt{2}} \left(A_2(\vec{x}+\vec{r}) A_3(\vec{x})
                     +A_3(\vec{x}+\vec{r}) A_2(\vec{x})\right),\\
\mathcal{O}_A^{T_2^{++},2}(r)&=\sum_{|\vec{r}|=r,\vec{x}}\frac{1}{\sqrt{2}} \left(A_1(\vec{x}+\vec{r}) A_3(\vec{x})
                     +A_3(\vec{x}+\vec{r}) A_1(\vec{x})\right),\\
\mathcal{O}_A^{T_2^{++},3}(r)&=\sum_{|\vec{r}|=r,\vec{x}}\frac{1}{\sqrt{2}}
\left(A_1(\vec{x}+\vec{r}) A_2(\vec{x})+A_2(\vec{x}+\vec{r}) A_1(\vec{x})\right),
\end{split}
\ee where the $\vec{r}$s with the same $|\vec{r}|=r$ and different spatial orientations are summed
up and the $t$ variable is omitted for convenience.

\section{Numerical details}
\label{numerical details} We generate the gauge configurations on two anisotropic lattices using
the tadpole-improved gauge action \cite{Morningstar:1997er}. The lattice sizes are $L^3\times
T=8^3\times 96$ and $L^3\times T=12^3\times 144$, respectively. The temporal lattice spacing $a_t$
is made much smaller than the spatial one $a_s$ with the anisotropy $\xi\equiv a_s/a_t=5$. Thus we
can obtain a higher resolution of hadron correlation functions in the temporal direction, which
helps us to handle glueball states whose signal-to-noise ratios damp rapidly in time. The relevant
input parameters are listed in Tab.~\ref{latticesetup}, where the $a_s$ values are determined
through the static potential with the scale parameter $r_0^{-1} = 410(20)\,{\rm MeV}$. There are
5000 gauge configurations generated for each lattice. The volumes of both lattices are roughly
$(1.7 \,{\rm fm})^3$, which have been tested to be large enough for glueballs~\cite{Chen:2006iy}.
\begin{table}[]
\caption{ The lattice parameters in this work, such as the coupling constant $\beta$, the
anisotropy $\xi$, and the number of the configurations, etc..  The ratio $a_s/r_0$ is determined by
the static potential. The first error of $a_s$ is statistical and the second one comes from the
uncertainty of the scale parameter $r_0^{-1}=410(20)$ MeV.\label{para}}
%\begin{ruledtabular}
  \begin{tabular}{ccccccc}
     $\beta$ &  $\xi$  & $a_s/r_0$ &$a_s$(fm) & $La_s$(fm)& $L^3\times T$ & $N_{\rm conf}$ \\\hline
       2.4  & 5  & 0.461(4) & 0.222(2)(11) & $\sim 1.78$ &$8^3\times 96$ & 5000 \\
      2.8  & 5  & 0.288(2) & 0.138(1)(7) & $\sim 1.66 $&$12^3\times 144$ & 5000  \\
    \end{tabular}
%  \end{ruledtabular}
\label{latticesetup}
\end{table}

The BS wave function in the rest frame of a glueball state can be derived by
calculating the following two-point functions of the two-gluon operator
$\mathcal{O}_A(t,r)$ defined above and an operator $\mathcal{O}_B$ which
creates glueball states with a specific quantum number,
\begin{eqnarray}\label{2pt2}
C_2(t,r)&=&\la 0 |\mathcal{O}_A(t,r) \mathcal{O}_B(0)|0 \ra\nonumber\\
&=&\sum\limits_{i}\frac{1}{2m_i}\la 0 |\mathcal{O}_A(r)|i\ra \la i| \mathcal{O}_B|0 \ra e^{-m_i t}\nonumber\\
&=&\sum\limits_{i}\Phi_i(r)e^{-m_i t}
\end{eqnarray}
where $m_i$ is the mass of the $i$-th state, and $\Phi_i(r)$ is its BS wave function by the
definition of Eq.~(\ref{wavefunc}) (up to an irrelevant normalization factor). Since the two-gluon
operators $O_A(t,r)$ are gauge variant objects, we need to calculate the two-point functions
$C_2(t,r)$ in a fixed gauge. In the practical calculation, all the gauge configurations are fixed
to the Coulomb gauge, in which the BS wave functions are usually assumed to have connection with
the wave functions of potential models.

In this work we intend to obtain the wave functions of both the ground and the first excited
glueball states, so the major task is to identify a specific state unambiguously. For this purpose,
we adopt the techniques applied in the calculation of the glueball spectrum to construct the
optimal glueball operators $\mathcal{O}_{\rm opt}$ which couple to specific states. Subsequently,
we use $\mathcal{O}_{\rm opt}$ as the operator $\mathcal{O}_B$ in Eq.~(\ref{2pt2}) to calculate the
required correlation functions. It should be noted that, with this prescription, the possible
mixing of glueballs to the flux states~\cite{deForcrand:1991kc} can be eliminated, or at least much
suppressed by the use of the optimal operators. We outline the construction of $\mathcal{O}_{\rm
opt}$ as follows.

As is mentioned above, the quantum numbers of glueballs are realized through the irreducible
representations $A_1$, $A_2$, $E$, $T_1$, and $T_2$ of the lattice point group $O$.  Therefore, we
first build the same 10 prototypes of Wilson loops as those in Refs.~\cite{Morningstar:1999cu,
Chen:2006iy}, based on which several smearing steps are applied to obtain more loop operators for
glueballs. We perform the 24 spatial operations of the $O$ group to these operators and linearly
combine them to realize the irreducible representations of $A_1$, $E$ and $T_2$. As such we obtain
24 different operators for each of the quantum numbers $R= A_1^{++}, E^{++}$, or $T_2^{++}$.
Finally, we implement the variational method to get the optimal operators by solving the
generalized eigenvalue problem. To be specific, for a given quantum number $R$, we first calculate
the $24\times 24$ correlation matrix $C^R_{\alpha\beta}(t)$ through the operator set
$\{\phi^{R}_\alpha, \alpha=1,2,\ldots,24\}$, \be C^R_{\alpha,\beta}(t)=\la 0| \phi^R_\alpha(t)
\phi^{R\dag}_\beta(0)|0 \ra, \ee and then solve the generalized eigenvalue problem \be C^R(t_0)
V^R_i = e^{-\tilde{m}^R_i t_0}C(0) V^R_i \ee
 to obtain the eigenvector $V^R_i$ for the $i$-th eigenvalue $e^{-\tilde{m}^R_i t_0}$.
 In this work we choose $t_0=1$. The eigenvector $V^R_i$ yields the combination
 coefficients for the optimal operator,
\be \mathcal{O}^R_{{\rm opt},i}=\sum_{\alpha=1}^{24} V^R_{i,\alpha}\cdot \phi^R_\alpha. \ee It has
been verified in the calculation of the spectrum that $\mathcal{O}^R_{{\rm opt},i} $ is highly
optimized and its correlation function is saturated by the $i$-th state even at the beginning time
slice. In other words, the correlation function of $\mathcal{O}^R_{{\rm opt},i}$ can be
consequently expressed as
\begin{equation}
C_{{\rm opt}}^{(R,i)}(t)=\la \mathcal{O}^R_{{\rm opt},i}(t)\mathcal{O}^{R\dag}_{{\rm opt},i}(0)\ra
\approx e^{-m_i^R t},
\end{equation}
where $C_{\rm opt}^{(R,i)}(t)$ is normalized as $C_{\rm opt}^{(R,i)}(0)=1$, and $m_i^R$ is the mass
of the state.

\begin{figure}[]
\includegraphics[width=0.45\textwidth]{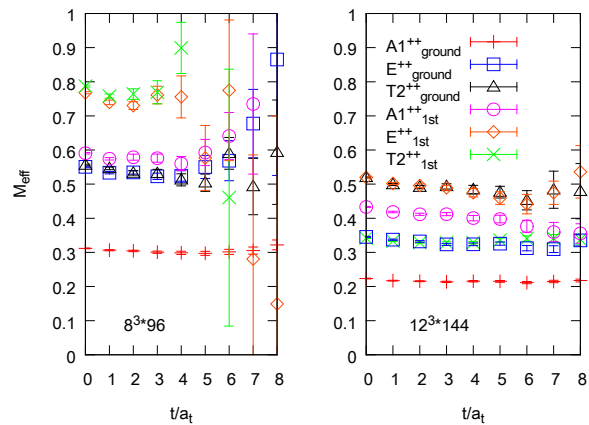}
\caption{Effective mass plateaus of the ground and the first excited states in
  $A_1^{++}, E^{++}$, and $T_2^{++}$ channels. The left plot is from
$8^3\times96$ lattice while the right one is from $12^3\times144$. The time and
the mass axis are both in lattice units.}
\label{mass_glb_12}
\end{figure}

\begin{table*}[]
\caption{ The masses are given in physical units for the ground and the first
  excited states of glueballs in the $A_1^{++}$, $E^{++}$, and $T_2^{++}$
  channels (the first excited states are labelled by an asterisk on the quantum
  numbers). The first error of each mass value is the statistical one and the
  second comes from the uncertainty of the scale parameter $r_0^{-1}=410(20)$
  MeV.\label{masses}}
\begin{ruledtabular}
  \begin{tabular}{ccccccc}
    $\beta$ &  $A_1^{++}$ (GeV) & $A_1^{++*}$ (GeV) & $E^{++}$ (GeV) &  $E^{++*}$ (GeV) & $T_2^{++}$ (GeV) &$T_2^{++*}$ (GeV)\\
  \hline
    2.4   &   1.36(1)(7)   &  2.58(4)(13)  & 2.40(4)(12) &  3.29(13)(16)   &  2.36(4)(12)  & 3.34(13)(16) \\
    2.8   &   1.52(2)(8)   &  2.92(20)(15) & 2.31(4)(11) &  3.42(15)(17)   &  2.31(3)(11)  & 3.49(21)(17) \\

    \end{tabular}
  \end{ruledtabular}
\end{table*}

The effective mass plateaus of the ground and the first excited states of scalar and tensor
glueballs are plotted in Fig.~\ref{mass_glb_12}, where one can see that the effective mass plateau
in each channel starts almost from the first time slice. The masses of the ground and the first
excited glueball states in the $A_1^{++}$, $E^{++}$, and $T_2^{++}$ channels are given in
Tab.~\ref{masses} in physical units, which are converted through the scale parameter
$r_0^{-1}=410(20)$ MeV. It is also seen that the masses of the $A_1^{++}$ glueballs have obvious
lattice spacing dependence, while the masses of the $E^{++}$ and $T_2^{++}$ glueballs are
insensitive to the lattice spacings. Our results are consistent with that of the previous
studies~\cite{Morningstar:1999cu, Chen:2006iy}. The coincidence of the masses of $E^{++}$ and
$T_2^{++}$ glueballs at the same lattice spacing implies that the effects due to the rotational
symmetry breaking are not important on the lattices we are using.

\begin{figure}[]
\includegraphics[width=0.45\textwidth]{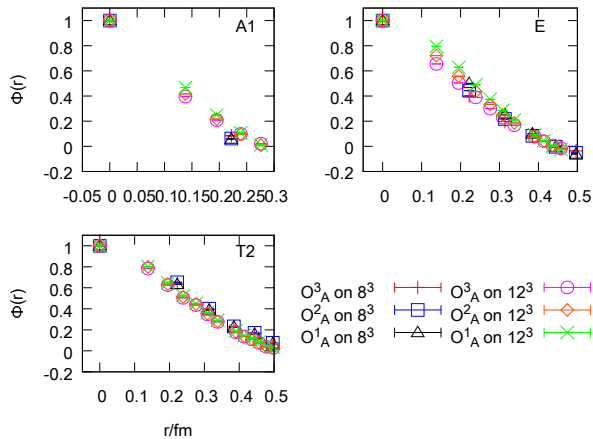}
\caption{The comparison of the ground state wave functions
  $\Phi(r)$ (normalized to 1 at $r=0$) extracted through the different
  definitions of the two-gluon operators in $A_1^{++}$, $E^{++}$, and
  $T_2^{++}$ channels. The finite lattice spacing effects are also checked by
  the comparison of the wave functions calculated on $8^3\times 96$ and
$12^3\times 144$ lattices. In each channel, the wave functions lie on each other within errors and
manifest the small lattice artifacts.} \label{plotwavefunction}
\end{figure}

By using the optimal glueball operators, the two-point function $C_2(t,r)$ can be simplified as \be
\label{2pt4} C_2^{(R,i)}(t,r)= \la 0 |\mathcal{O}^R_A(t,r) \mathcal{O}^{R\dag}_{{\rm opt},i}(0)|0
\ra \approx N ~\Phi_i^R(r)~ e^{-m_i^R t}. \ee where $m_i^R$ is the mass of the glueball state in
concern, $N$ is an irrelevant normalization factor, and $\Phi_i^R(r)$ is the desired wave function.
Consequently, $\Phi_R(r)$ can be directly obtained through a single-exponential fit. For a definite
state, since the mass $m^R$ is a common parameter of the two-point functions $C_2^{(R,i)}(t,r)$
with different $r$s as well as the relevant $C_{\rm opt}^{(R,i)}(t)$, we perform a joint data
analysis on these correlation functions. We first divide the 5000 measurements of the two-point
functions into 100 bins and take the average of the 50 measurements in each bin as an independent
measurement. After that we construct a bootstrapped covariance matrix of these correlation and
carry out a correlated minimal-$\chi^2$ data fitting, through which the wave functions
$\Phi_i^R(r)$ at different $r$s can be obtained simultaneously . For all the channels we are
interested in, the fitting qualities are very good.

We first test the three definitions of the two-gluon operators on the two lattices we are using.
The ground state wave functions in the $A_1^{++}$, $E^{++}$, and $T_2^{++}$ channels are plotted in
Fig.~\ref{plotwavefunction} for comparison, where the spatial distances $r$ are expressed in
physical units based on the lattice spacings listed in Tab.~\ref{para}. In each channel, the wave
functions from the three types of the two-gluon operators lie on the same curve within errors. This
implies that the three definitions of the two-gluon operators are numerically equivalent. On the
other hand, the wave functions from the two lattices do not show sizable difference and manifest
that the finite $a$ artifacts are not important. This observation is understandable since we are
using the improved action, which suppresses the discretization error substantially. Therefore, in
the following data analysis and discussions in this paper, we focus only on the wave functions
derived from the finer lattice with the third definition of the two-gluon operators.

\begin{table}[]
  \caption{The best-fit parameters of the ground state wave functions using the parameterization
  in Eq.~(\ref{ground}). The RMS radii $r_{\rm RMS}$ of ground state glueball wave functions are also
  listed. The size of the tensor glueball is roughly twice as large as that of the scalar glueball.}
  \begin{ruledtabular}
  \begin{tabular}{cccc}
           & $A_1^{++}$ & $E^{++}$ & $T_2^{++}$\\
       \hline
     $r_0$ (fm) & $0.154(8)$ & $0.258(5)$ & $0.296(3)$\\
     $\alpha$   & $2.1(4)$ & $ 2.1(1) $ & $2.07(7)$\\
      $r_{\rm RMS}$ (fm)   & $0.127$ & $ 0.211 $ & $0.248$
    \end{tabular}
  \end{ruledtabular}
\label{fitresult}
\end{table}

\begin{figure}[!h]
\includegraphics[width=0.45\textwidth]{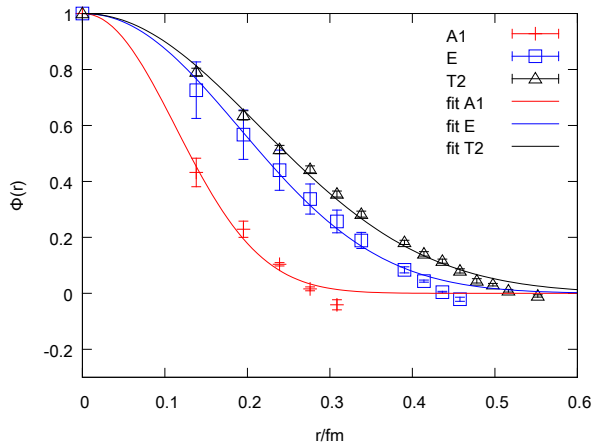}
\caption{The ground state wave functions derived on the $12^3\times 144$ lattice
  are plotted by data points. The curves are the fitting results using the
  parameterization in Eq.~(\ref{ground}).}
\label{fitplot2}
\end{figure}

In order to get some quantitative information, we parameterize the ground state wave functions as
the following function form, \be\label{ground}
\Phi_1(r)=\Phi_1(0)e^{-\left(\frac{r}{r_0}\right)^\alpha}, \ee where the parameters $r_0$ and
$\alpha$ can be determined by fitting the lattice data. Since the measured values of the wave
functions at different $r$ are correlated, in extracting $r_0$ and $\alpha$ we perform the
correlated data fitting with a bootstrapped covariance matrix.  The best-fit results are listed in
Tab.~\ref{fitresult} for the three channels on the $12^3$ lattice. It is interesting that the
parameter $\alpha$ is close to 2 within errors for all the cases. This is very different from the
previous work~\cite{deForcrand:1991kc,Loan:2006hw} where $\Phi_1(r)$ is Coulomb type instead of the
Gaussian type in this study. The fitted wave functions of the ground states for the three channels
are plotted in Fig.~\ref{fitplot2} for illustration. It should be noted that the wave functions of
$E^{++}$ and $T_2^{++}$ ground state glueballs are slightly different, which can be attributed to
the rotational symmetry breaking on the lattice. Obviously, the effect of this kind of symmetry
breaking is enlarged in the spatially extended quantities such as the wave functions, even though
their masses are nearly degenerate.

We also calculate the root-of-mean-square (RMS) radii $r_{\rm RMS}$ for the
ground state glueballs with the definition
\begin{equation}
r_{\rm RMS}^2=\frac{\int dr r^4 \Phi^2(r)}{\int dr r^2 \Phi^2(r)},
\end{equation}
where $\Phi(r)$ is the wave function with the best-fit parameters. The results are also listed in
Tab.~\ref{fitresult}.  The $r_{\rm RMS}$'s of the $E^{++}$ and $T_2^{++}$ glueballs are almost
twice as large as that of the $A_1^{++}$ glueball.

\begin{figure}[]
\includegraphics[width=0.45\textwidth]{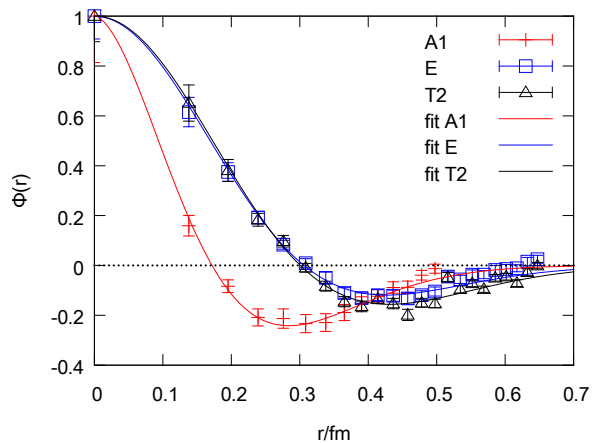}
\caption{The wave functions of the first excited states. The data points are measured results and
the curves illustrate the parameterization in Eq.~(\ref{excitedwave}) with the best-fit parameters.
} \label{excited}
\end{figure}

\begin{table}[!h]
  \caption{The best-fit results of wave functions of the first excited states
    in the three channels using the parameterization of
    Eq.~(\ref{excitedwave}). The errors are only statistical.}
  \begin{ruledtabular}
  \begin{tabular}{cccc}
           & $A_1^{++}$ & $E^{++}$ & $T_2^{++}$\\
       \hline
     $\beta$ & $21(4)$ & $11(1)$ & $11(1)$\\
     $\alpha$   & $1.7(1)$ & $ 2.0(1) $ & $2.0(1)$\\
      $r_0$(fm)   & $0.21(1)$ & $ 0.30(1) $ & $0.31(1)$
    \end{tabular}
  \end{ruledtabular}
\label{fitresult2}
\end{table}

Benefiting from the variational method discussed above, we can also identify clearly the first
excited states of glueballs in the three channels. This enables us to derive their wave functions
as well, and the procedure is similar to that for the ground states. Fig.~\ref{excited} (the
points) shows the wave functions of the $A_1^{++}$ , $E^{++}$ and $T_2^{++}$ excited glueballs. We
do observe the radial nodes for all these wave functions, which can be viewed as a strong evidence
that the excited glueball is the first radial excited state of the ground glueball in each channel.
Inspired by the function form of the two-body non-relativistic Schr\"{o}dinger equation, we
parameterize the wave functions of the first excited states as follows, \be \label{excitedwave}
\Phi_2(r)=\Phi_2(0)(1-\beta\cdot r^\alpha)e^{-\left(\frac{r}{r_0}\right)^\alpha}, \ee where the
parameter $\alpha$ depends on the type of the interacting potential (for example, $\alpha=1$ for
the Coulomb potential and $\alpha=2$ for the harmonious oscillator potential). The parameters
$r_0$, $\alpha$, and $\beta$ are determined through a procedure similar to that for the ground
states, and the best-fit results for the $12^3\times 144$ lattice are listed in
Tab.~\ref{fitresult2}. It is interesting that $\alpha$ and $r_0$ in the tensor channel ($T_2^{++}$
and $E^{++}$) are close to those of the ground state. Especially, $\alpha$ is still close to 2,
which resembles the case for the harmonious oscillator potential model. For the first excited state
of the scalar glueball, the fitted parameter $\alpha$ deviates from 2 a little. The reason for this
is not clear yet. Recalling that the size of ground scalar glueball is as small as 0.127 fm, while
the lattice spacings we are using are comparable to it or even larger, so a possibility is that the
radial behavior of the first excited scalar glueball is not extracted accurately enough.

%Based on the observation of the BS wave functions of glueballs in this study, one can infer some
%helpful information on the inner structure of glueballs. For the ground states, the BS wave
%functions of the scalar and the tensor glueball show the similar radial behaviors which can be well
%described by Gaussian-like function forms. Assuming dominant two-gluon components in these glueball
%states, the lowest glueball state must be the scalar glueball, whose two constituent gluons are in
%$S$-wave. Furthermore, the similarity of the ground state wave functions also implies that the two
%constituent gluons of the tensor glueballs are in the $S$-wave. With the parameterizations in
%Eq.~(\ref{ground}) and Eq.~(\ref{excitedwave}), the wave functions of the ground and the excited
%tensor glueballs are compatible with the $1S$ and $2S$ wave functions of a non-relativistic
%harmonious oscillator. This may provide useful information for understanding the interaction
%between gluons within glueballs in the potential model picture.

\section{ Summary and Discussions}\label{summary}
The Bethe-Salpeter wave functions of the pure gauge scalar and tensor
glueballs are revisited in this work. The feature of this study is the
precise identification of the ground and the first excited glueball
states in these two channels through the variational method based on
large operator sets. We test the different definitions of two-gluon
operators and find no sizable difference, which implies the finite
lattice spacing artifacts are small due to the implementation of the
improved gauge action. This is also reinforced by checking the results
from two lattices with different lattice spacings.

With large statistics, the BS wave functions of both the ground and the first excited states in the
scalar and tensor channels are extracted precisely.  Instead of the exponential fall-off of the
wave functions observed in previous works, we find that the wave functions of the ground states are
Gaussian-like.  The size of the ground state tensor glueball is roughly twice as large as that of
the ground state scalar glueball. For the first time, we observe the radial nodes of wave functions
of the first excited states, which support them as the first radial excitations. We use the
function forms inspired by potential models to parameterize the wave functions, and the fitted
parameters show that the wave functions of the ground and the first excited state in the tensor
channel are compatible with the $1S$ and $2S$ wave functions of a harmonious oscillator.  These
observations are helpful for a qualitative understanding of the inner structure of glueballs.
Furthermore, there have been quite a few phenomenological studies on glueballs in the picture of
potential models~\cite{Mathieu:2008es,Buisseret:2009dk,Buisseret:2007cx,Boulanger:2008dt}, which
use the scalar glueball wave function as an input to derive the interacting potential between
constituent gluons. The results in this work can provide much finer and more precise information on
this sector.

\section*{ACKNOWLEDGEMENTS}
This work is supported in part by the National Science Foundation of China
(NSFC) under Grants No. 10835002, No. 11075167, No. 11105153, No. 11335001, and
11405053. Z.L. is partially supported by the Youth Innovation Promotion
Association of CAS. Y.C. and Z.L. also acknowledge the support of NSFC under
No. 11261130311 (CRC 110 by DFG and NSFC)

%\bibliographystyle{unsrt}%ordered as cited
%\bibliographystyle{apsrev4-1}
%\bibliography{bib/tex}

\begin{thebibliography}{10}

\bibitem{Morningstar:1997er} C. Morningstar and M. Peardon, Phys. Rev. D {\bf 56}, 4043 (1997).
\bibitem{Morningstar:1999cu} C. Morningstar and M. Peardon, Phys. Rev. D {\bf 60}, 034509 (1999).
\bibitem{Chen:2006iy} Y. Chen {\it et al.}, Phys. Rev. D {\bf 73}, 014516 (2006).
\bibitem{Gregory:2012cr} E. Gregory, A. Irving, B. Lucini, C. McNeile, A. Rago, C. Richards, and E. Rinaldi, J. High Energy Phys. {\bf 10} (2012) 170.
\bibitem{Carlson:1983} C.E. Carlson, T.H. Hansson, C. Peterson, Phys. Rev. D {\bf 27}, 1556 (1983).
\bibitem{Szczepaniak:1996} A. Szczepaniak, E.S. Swanson, C.-R. Ji, S.R. Cotanch, Phys. Rev. Lett. {\bf 76}, 2011(1996) [arXiv:hep-ph/9511422].
\bibitem{Szczepaniak:2003} A. Szczepaniak, E.S. Swanson, Phys. Lett. B {\bf 577}, 61 (2003) [arXiv:hep-ph/0308268].
\bibitem{Fritzsch:1981} H. Fritzsch, P. Minkowski, Nuovo Cimento A {\bf 30}, 393 (1975).
\bibitem{Barnes:1981} T. Barnes, Z.Phys. C {\bf 10}, 275 (1981).
\bibitem{Cornwall:1983} J.M. Cornwall, A. Soni, Phys. Lett. B {\bf 120}, 431 (1983).
\bibitem{deForcrand:1991kc} P. de Forcrand and K.-F. Liu, Phys. Rev. Lett. {\bf 69}, 245 (1992).
\bibitem{Loan:2006hw} M. Loan and Y. Ying, Prog. Theor. Phys. {\bf 116}, 169 (2006).
\bibitem{Buisseret:2007cx} F. Buisseret and C. Semay, Eur. Phys. J. A {\bf 33}, 87 (2007).
\bibitem{Buisseret:2009dk} F. Buisseret, Phys. Rev. D {\bf 79}, 037503 (2009).
\bibitem{Salpeter:1951} E.E. Salpeter, H.A. Bethe, Phys. Rev. {\bf 84}, 1232 91951); J. Schwinger, Proc. Natl. Acad. Sci. U.S.A. {\bf 37}, 455 (1951).
\bibitem{Boulanger:2008dt} N. Boulanger, F. Buisseret, V. Mathieu, and C. Semay, Eur. Phys. J. A {\bf 38}, 317 (2008).
\bibitem{Mathieu:2008es} V. Mathieu, F. Buisseret, and C. Semay, Phys. Rev. D {\bf 77}, 114022 (2008).
\end{thebibliography}
%\nocite{*}

\end{document}